\documentclass[11pt]{article}
\pdfoutput=1
\usepackage{graphicx}
\usepackage{amssymb}
\usepackage{amsmath,bm}
\usepackage[left]{lineno}
\usepackage{subfigure}
\usepackage{booktabs}
\usepackage{multirow}
\usepackage[title]{appendix}
\usepackage{lineno}
\usepackage[T1]{fontenc}
\usepackage[utf8]{inputenc}
\usepackage{booktabs,siunitx}
\usepackage{verbatim}
\usepackage[tablewithin=none,figurewithin=none,singlelinecheck=false]{caption} 
\usepackage{longtable,booktabs}
\usepackage{lineno}

\usepackage{longtable}

\def\kpp{K^+\rightarrow\pi^+\pi^0}

\def\pgee{\pi^0\rightarrow\gamma e^+ e^-}
\def\pgnn{\pi^0\rightarrow\gamma\nu\bar{\nu}}
\def\kenp{K^+\rightarrow\pi^0 e^+\nu_e}
\def\kmnp{K^+\rightarrow\pi^0 \mu^+\nu_\mu}
\def\kppgDE{K^+\rightarrow\pi^+\pi^0\gamma (\mathrm{DE})}
\def\kppgINT{K^+\rightarrow\pi^+\pi^0\gamma (\mathrm{INT})}
\def\kmnug{K^+\rightarrow\mu^+\nu_\mu(\gamma)}
\def\kenug{K^+\rightarrow e^+\nu_e(\gamma)}
\def\kppp{K^+\rightarrow\pi^+\pi^0\pi^0}

\begin{document}
\pagenumbering{arabic}
\centerline{\LARGE EUROPEAN ORGANIZATION FOR NUCLEAR RESEARCH}

\begin{flushright}
CERN-EP-2019-048  \\
March 26, 2019\\
Revised version\\
\today \\
\end{flushright}
\vspace{15mm}

\begin{center}
\Large{\bf 
Search for production of an invisible dark photon in $\mathbf{\pi^0}$ decays\\
\vspace{5mm}
}
The NA62 Collaboration
\end{center}
\vspace{10mm}

\begin{abstract}
The results of a search for $\pi^0$ decays to a photon and an invisible massive dark photon at the NA62 experiment at the CERN SPS are reported. From a total of $4.12\times10^{8}$ tagged $\pi^0$ mesons, no signal is observed. Assuming a kinetic-mixing interaction, limits are set on the dark photon coupling to the ordinary photon as a function of the dark photon mass, improving on previous searches in the mass range 60--110~MeV$/c^2$. The present results are interpreted in terms of an upper limit of the branching ratio of the electro-weak decay $\pi^0 \to \gamma \nu \bar\nu$, improving the current limit by more than three orders of magnitude.
\end{abstract}
\vspace{10mm}
\clearpage
\begin{center}
{\Large The NA62 Collaboration$\,$\renewcommand{\thefootnote}{\fnsymbol{footnote}}%
\footnotemark[1]\renewcommand{\thefootnote}{\arabic{footnote}}}\\
\end{center}
\vspace{2mm}
\begin{raggedright}
\noindent
{\bf Universit\'e Catholique de Louvain, Louvain-La-Neuve, Belgium}\\
 E.~Cortina Gil,
 A.~Kleimenova,
 E.~Minucci$\,$\footnotemark[1],
 S.~Padolski$\,$\footnotemark[2],
 P.~Petrov,
 A.~Shaikhiev$\,$\footnotemark[3],
 R.~Volpe\\[2mm]

{\bf Faculty of Physics, University of Sofia, Sofia, Bulgaria}\\
 G.~Georgiev$\,$\footnotemark[4],
 V.~Kozhuharov$\,$\footnotemark[4],
 L.~Litov\\[2mm]

{\bf TRIUMF, Vancouver, British Columbia, Canada}\\
 T.~Numao,
 Y.~Petrov,
 B.~Velghe\\[2mm]

{\bf University of British Columbia, Vancouver, British Columbia, Canada}\\
 D.~Bryman,
 J.~Fu$\,$\footnotemark[5]\\[2mm]

{\bf Charles University, Prague, Czech Republic}\\
 T.~Husek$\,$\footnotemark[6],
 J.~Jerhot,
 K.~Kampf,
 M.~Zamkovsky\\[2mm]

{\bf Institut f\"ur Physik and PRISMA Cluster of excellence, Universit\"at Mainz, Mainz, Germany}\\
 R.~Aliberti,
 G.~Khoriauli$\,$\footnotemark[7],
 J.~Kunze,
 D.~Lomidze$\,$\footnotemark[8],
 L.~Peruzzo,
 M.~Vormstein,
 R.~Wanke\\[2mm]

{\bf Dipartimento di Fisica e Scienze della Terra dell'Universit\`a e INFN, Sezione di Ferrara, Ferrara, Italy}\\
 P.~Dalpiaz,
 M.~Fiorini,
 I.~Neri,
 A.~Norton,
 F.~Petrucci,
 H.~Wahl\\[2mm]

{\bf INFN, Sezione di Ferrara, Ferrara, Italy}\\
 A.~Cotta Ramusino,
 A.~Gianoli\\[2mm]

{\bf Dipartimento di Fisica e Astronomia dell'Universit\`a e INFN, Sezione di Firenze, Sesto Fiorentino, Italy}\\
 E.~Iacopini,
 G.~Latino,
 M.~Lenti,
 A.~Parenti\\[2mm]

{\bf INFN, Sezione di Firenze, Sesto Fiorentino, Italy}\\
 A.~Bizzeti$\,$\footnotemark[9],
 F.~Bucci\\[2mm]

{\bf Laboratori Nazionali di Frascati, Frascati, Italy}\\
 A.~Antonelli,
 G.~Lanfranchi,
 G.~Mannocchi,
 S.~Martellotti,
 M.~Moulson,
 T.~Spadaro\renewcommand{\thefootnote}{\fnsymbol{footnote}}%
\footnotemark[1]\renewcommand{\thefootnote}{\arabic{footnote}}\\[2mm]

{\bf Dipartimento di Fisica ``Ettore Pancini'' e INFN, Sezione di Napoli, Napoli, Italy}\\
 F.~Ambrosino,
 T.~Capussela,
 M.~Corvino,
 D.~Di Filippo,
 P.~Massarotti,
 M.~Mirra\renewcommand{\thefootnote}{\fnsymbol{footnote}}%
\footnotemark[1]\renewcommand{\thefootnote}{\arabic{footnote}},
 M.~Napolitano,
 G.~Saracino\\[2mm]

{\bf Dipartimento di Fisica e Geologia dell'Universit\`a e INFN, Sezione di Perugia, Perugia, Italy}\\
 G.~Anzivino,
 F.~Brizioli,
 E.~Imbergamo,
 R.~Lollini,
 R.~Piandani,
 C.~Santoni\\[2mm]

{\bf INFN, Sezione di Perugia, Perugia, Italy}\\
 M.~Barbanera$\,$\footnotemark[10],
 P.~Cenci,
 B.~Checcucci,
 P.~Lubrano,
 M.~Lupi$\,$\footnotemark[11],
 M.~Pepe,
 M.~Piccini\\[2mm]

{\bf Dipartimento di Fisica dell'Universit\`a e INFN, Sezione di Pisa, Pisa, Italy}\\
 F.~Costantini,
 L.~Di Lella,
 N.~Doble,
 M.~Giorgi,
 S.~Giudici,
 G.~Lamanna,
 E.~Lari,
 E.~Pedreschi,
 M.~Sozzi\\[2mm]

{\bf INFN, Sezione di Pisa, Pisa, Italy}\\
 C.~Cerri,
 R.~Fantechi,
 L.~Pontisso,
 F.~Spinella\\[2mm]

{\bf Scuola Normale Superiore e INFN, Sezione di Pisa, Pisa, Italy}\\
 I.~Mannelli\\[2mm]

{\bf Dipartimento di Fisica, Sapienza Universit\`a di Roma e INFN, Sezione di Roma I, Roma, Italy}\\
 G.~D'Agostini,
 M.~Raggi\\[2mm]

{\bf INFN, Sezione di Roma I, Roma, Italy}\\
 A.~Biagioni,
 E.~Leonardi,
 A.~Lonardo,
 P.~Valente,
 P.~Vicini\\[2mm]

{\bf INFN, Sezione di Roma Tor Vergata, Roma, Italy}\\
 R.~Ammendola,
 V.~Bonaiuto$\,$\footnotemark[12],
 A.~Fucci,
 A.~Salamon,
 F.~Sargeni$\,$\footnotemark[13]\\[2mm]

{\bf Dipartimento di Fisica dell'Universit\`a e INFN, Sezione di Torino, Torino, Italy}\\
 R.~Arcidiacono$\,$\footnotemark[14],
 B.~Bloch-Devaux,
 M.~Boretto$\,$\footnotemark[15],
 E.~Menichetti,
 E.~Migliore,
 D.~Soldi\\[2mm]

{\bf INFN, Sezione di Torino, Torino, Italy}\\
 C.~Biino,
 A.~Filippi,
 F.~Marchetto\\[2mm]

{\bf Instituto de F\'isica, Universidad Aut\'onoma de San Luis Potos\'i, San Luis Potos\'i, Mexico}\\
 J.~Engelfried,
 N.~Estrada-Tristan$\,$\footnotemark[16]\\[2mm]

{\bf Horia Hulubei national Institute of Physics and Nuclear Engineering, Bucharest-Magurele, Romania}\\
 A. M.~Bragadireanu,
 S. A.~Ghinescu,
 O. E.~Hutanu\\[2mm]

{\bf Joint Institute for Nuclear Research, Dubna, Russia}\\
 T.~Enik,
 V.~Falaleev,
 V.~Kekelidze,
 A.~Korotkova,
 D.~Madigozhin,
 M.~Misheva$\,$\footnotemark[17],
 N.~Molokanova,
 S.~Movchan,
 I.~Polenkevich,
 Yu.~Potrebenikov,
 S.~Shkarovskiy,
 A.~Zinchenko$\,$\renewcommand{\thefootnote}{\fnsymbol{footnote}}\footnotemark[2]\renewcommand{\thefootnote}{\arabic{footnote}}\\[2mm]

{\bf Institute for Nuclear Research of the Russian Academy of Sciences, Moscow, Russia}\\
 S.~Fedotov,
 E.~Gushchin,
 A.~Khotyantsev,
 Y.~Kudenko$\,$\footnotemark[18],
 V.~Kurochka,
 M.~Medvedeva,
 A.~Mefodev\\[2mm]

{\bf Institute for High Energy Physics - State Research Center of Russian Federation, Protvino, Russia}\\
 S.~Kholodenko,
 V.~Kurshetsov,
 V.~Obraztsov,
 A.~Ostankov,
 V.~Semenov$\,$\renewcommand{\thefootnote}{\fnsymbol{footnote}}\footnotemark[2]\renewcommand{\thefootnote}{\arabic{footnote}},
 V.~Sugonyaev,
 O.~Yushchenko\\[2mm]

{\bf Faculty of Mathematics, Physics and Informatics, Comenius University, Bratislava, Slovakia}\\
 L.~Bician,
 T.~Blazek,
 V.~Cerny,
 Z.~Kucerova\\[2mm]

{\bf CERN,  European Organization for Nuclear Research, Geneva, Switzerland}\\
 J.~Bernhard,
 A.~Ceccucci,
 H.~Danielsson,
 N.~De Simone$\,$\footnotemark[19],
 F.~Duval,
 B.~D\"obrich,
 L.~Federici,
 E.~Gamberini,
 L.~Gatignon,
 R.~Guida,
 F.~Hahn$\,$\renewcommand{\thefootnote}{\fnsymbol{footnote}}\footnotemark[2]\renewcommand{\thefootnote}{\arabic{footnote}},
 E. B.~Holzer,
 B.~Jenninger,
 M.~Koval,
 P.~Laycock$\,$\footnotemark[2],
 G.~Lehmann Miotto,
 P.~Lichard,
 A.~Mapelli,
 R.~Marchevski,
 K.~Massri,
 M.~Noy,
 V.~Palladino$\,$\footnotemark[20],
 M.~Perrin-Terrin$\,$\footnotemark[21]$^,\,$\footnotemark[22],
 J.~Pinzino,
 V.~Ryjov,
 S.~Schuchmann,
 S.~Venditti\\[2mm]

{\bf University of Birmingham, Birmingham, United Kingdom}\\
 T.~Bache,
 M. B.~Brunetti,
 V.~Duk,
 V.~Fascianelli$\,$\footnotemark[23],
 J. R.~Fry,
 F.~Gonnella,
 E.~Goudzovski,
 L.~Iacobuzio,
 C.~Lazzeroni,
 N.~Lurkin,
 F.~Newson,
 C.~Parkinson,
 A.~Romano,
 A.~Sergi,
 A.~Sturgess,
 J.~Swallow\\[2mm]

{\bf University of Bristol, Bristol, United Kingdom}\\
 H.~Heath,
 R.~Page,
 S.~Trilov\\[2mm]

{\bf University of Glasgow, Glasgow, United Kingdom}\\
 B.~Angelucci,
 D.~Britton,
 C.~Graham,
 D.~Protopopescu\\[2mm]

{\bf University of Lancaster, Lancaster, United Kingdom}\\
 J.~Carmignani,
 J. B.~Dainton,
 R. W. L.~Jones,
 G.~Ruggiero\\[2mm]

{\bf University of Liverpool, Liverpool, United Kingdom}\\
 L.~Fulton,
 D.~Hutchcroft,
 E.~Maurice$\,$\footnotemark[24],
 B.~Wrona\\[2mm]

{\bf George Mason University, Fairfax, Virginia, USA}\\
 A.~Conovaloff,
 P.~Cooper,
 D.~Coward$\,$\footnotemark[25],
 P.~Rubin\\[2mm]

\end{raggedright}
%
%
\setcounter{footnote}{0}
\renewcommand{\thefootnote}{\fnsymbol{footnote}}
\footnotetext[1]{Corresponding authors: T.~Spadaro, M.~Mirra \\ 
email: tommaso.spadaro@cern.ch, marco.mirra@cern.ch}
\footnotetext[2]{Deceased}
\renewcommand{\thefootnote}{\arabic{footnote}}

\footnotetext[1]{Present address: Laboratori Nazionali di Frascati, I-00044 Frascati, Italy}
\footnotetext[2]{Present address: Brookhaven National Laboratory, Upton, NY 11973, USA}
\footnotetext[3]{Also at Institute for Nuclear Research of the Russian Academy of Sciences, 117312 Moscow, Russia}
\footnotetext[4]{Also at Laboratori Nazionali di Frascati, I-00044 Frascati, Italy}
\footnotetext[5]{Present address: UCLA Physics and Biology in Medicine, Los Angeles, CA 90095, USA}
\footnotetext[6]{Present address: IFIC, Universitat de Val\`encia - CSIC, E-46071 Val\`encia, Spain}
\footnotetext[7]{Present address: Universit\"at W\"urzburg, D-97070 W\"urzburg, Germany}
\footnotetext[8]{Present address: Universit\"at Hamburg, D-20146 Hamburg, Germany}
\footnotetext[9]{Also at Dipartimento di Fisica, Universit\`a di Modena e Reggio Emilia, I-41125 Modena, Italy}
\footnotetext[10]{Present address: INFN, Sezione di Pisa, I-56100 Pisa, Italy}
\footnotetext[11]{Present address: Institut am Fachbereich Informatik und Mathematik, Goethe Universit\"at, D-60323 Frankfurt am Main, Germany}
\footnotetext[12]{Also at Department of Industrial Engineering, University of Roma Tor Vergata, I-00173 Roma, Italy}
\footnotetext[13]{Also at Department of Electronic Engineering, University of Roma Tor Vergata, I-00173 Roma, Italy}
\footnotetext[14]{Also at Universit\`a degli Studi del Piemonte Orientale, I-13100 Vercelli, Italy}
\footnotetext[15]{Also at CERN,  European Organization for Nuclear Research, CH-1211 Geneva 23, Switzerland}
\footnotetext[16]{Also at Universidad de Guanajuato, Guanajuato, Mexico}
\footnotetext[17]{Present address: Institute of Nuclear Research and Nuclear Energy of Bulgarian Academy of Science (INRNE-BAS), BG-1784 Sofia, Bulgaria}
\footnotetext[18]{Also at National Research Nuclear University (MEPhI), 115409 Moscow and Moscow Institute of Physics and Technology, 141701 Moscow region, Moscow, Russia}
\footnotetext[19]{Present address: DESY, D-15738 Zeuthen, Germany}
\footnotetext[20]{Present address: Physics Department, Imperial College London, London, SW7 2BW, UK}
\footnotetext[21]{Present address: Centre de Physique des Particules de Marseille, Universit\'e Aix Marseille, CNRS/IN2P3, F-13288, Marseille, France}
\footnotetext[22]{Also at Universit\'e Catholique de Louvain, B-1348 Louvain-La-Neuve, Belgium}
\footnotetext[23]{Present address: Dipartimento di Psicologia, Universit\`a di Roma La Sapienza, I-00185 Roma, Italy}
\footnotetext[24]{Present address: Laboratoire Leprince Ringuet, F-91120 Palaiseau, France}
\footnotetext[25]{Also at SLAC National Accelerator Laboratory, Stanford University, Menlo Park, CA 94025, USA}

\clearpage
\section{Introduction}
\label{sec:intro}
One of the possible extensions of the Standard Model (SM) aimed at explaining the abundance of dark matter in our universe predicts a new $U(1)$ gauge-symmetry sector with a vector mediator field $A^\prime$, often called ``dark photon''. In a simple realization of such a scenario~\cite{Okun,Holdom}, an $A^\prime$ field $A^\prime_{\mu\nu}$ with mass $M_{A^\prime}$ interacts with the SM photon through a kinetic-mixing Lagrangian, 
\begin{equation}
\label{eq:mixing}
\epsilon A^\prime_{\mu\nu}F^{\mu\nu},
\end{equation}
where $F_{\mu\nu}$ represents the electromagnetic field tensor and $\epsilon<<1$ is the coupling constant. A consequence of this interaction is the transition $\pi^0\to A^\prime\gamma$ with branching ratio, BR:
\begin{equation}
\label{eq:br}
\mathrm{BR}\left(\pi^0 \to A^\prime \gamma\right) = 2\epsilon^2\left(1 - \frac{M_{A^\prime}^2}{M_{\pi^{0}}^2}\right)^3\times\mathrm{BR}\left(\pi^0 \to \gamma \gamma\right).
\end{equation}
In a general picture, the above Lagrangian might be accompanied by further interactions, both with SM matter fields and with a 
secluded hidden sector of possible dark-matter candidate fields. If these are lighter than the $A^\prime$, the dark photon would decay mostly invisibly.

The search for an invisible $A^\prime$ is performed with a missing-mass technique from the full reconstruction of the decay chain
\begin{equation}
\label{eq:decaychain}
 K^+ \to \pi^+\pi^0,\quad \pi^0 \to A^\prime \gamma.
\end{equation}
An abundant flux of $K^+$ mesons is provided by a high-energy unseparated hadron beam from the CERN Super Proton Synchrotron (SPS). The search is performed using the NA62 experiment, which has the main goal of measuring the BR of the rare decay $K^+ \rightarrow \pi^+ \nu \bar\nu$ with 10\% precision. 
The design of the experiment guarantees high intensity, full particle identification, hermetic coverage, low material budget and high-rate tracking. 
The NA62 detector has been fully operational since 2016. The results from the analysis of a subsample of 2016 data are reported, corresponding to $1\%$ of the statistics collected by NA62 in 2016--2018.

\section{Beam line and detector}

\begin{figure}[!ht]
\centerline{\includegraphics[width=1.0\linewidth]{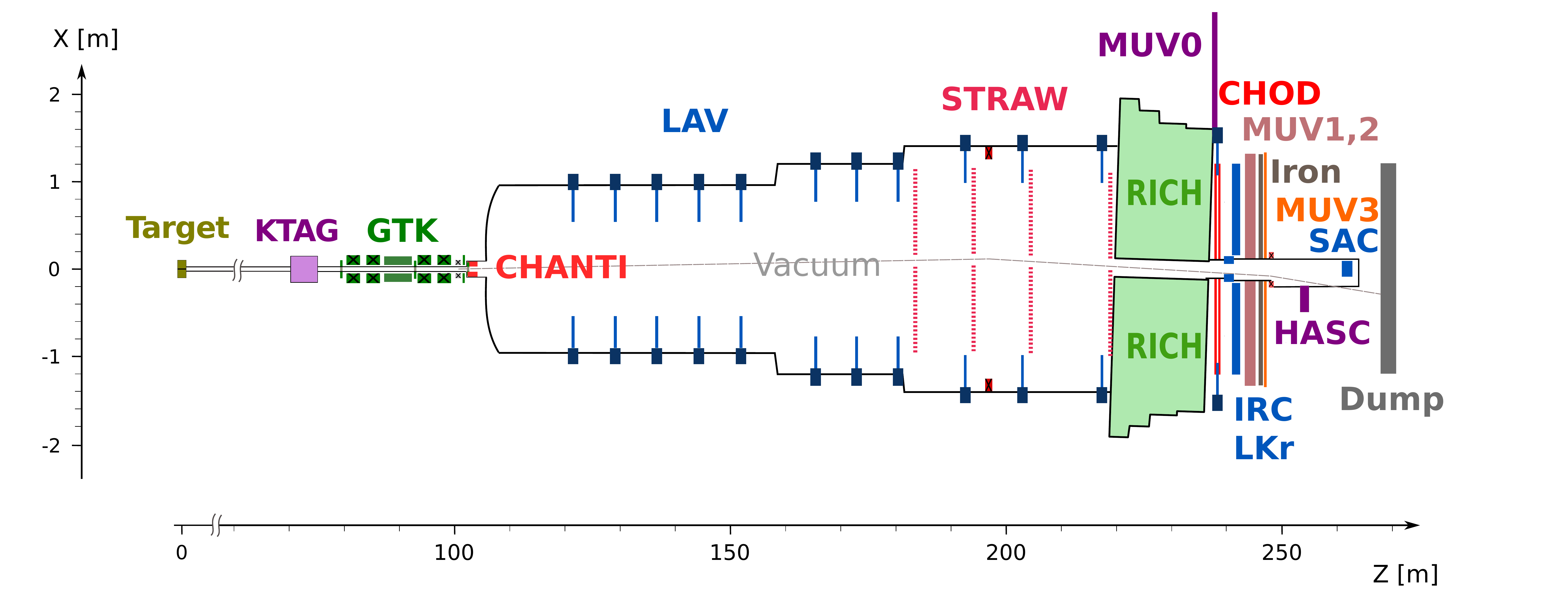}}
\hfill
 \caption{Schematic top view of the NA62 beam line and detector. The trajectory of a beam particle is shown, travelling in vacuum and crossing the detector apertures. A dipole magnet between the MUV3 and SAC systems deflects beam particles out of the SAC acceptance.}
\label{fig:NA62Apparatus}
\end{figure}
The beam line and detector, shown schematically in Fig.~\ref{fig:NA62Apparatus}, are described in detail elsewhere~\cite{NA62paper}. Here, the aspects relevant to the search for the decay chain described in Eq.~\ref{eq:decaychain} are outlined.

A proton beam of 400 GeV$/c$ in a 4.8 s long spill from the SPS hits a beryllium target to produce an intense 75 GeV$/c$ secondary beam of  positive particles, 6\% of which are charged kaons. The secondary beam is selected with a 1\% RMS momentum bite and is transported to the decay region more than 100~m downstream of the target. For the present measurement, the typical beam particle rate at the entrance of the decay volume is 300~MHz. Incoming kaons are positively identified by a differential Cherenkov counter read out by photomultipliers (PMs) grouped into eight sectors (KTAG): requiring a set of in-time signals (KTAG candidate) in five or more sectors identifies a $K^+$ with 70~ps time resolution. A magnetic spectrometer hosting three stations of Si-pixel detectors (GTK) reconstructs tracks for individual beam particles with 100~ps resolution and provides the longitudinal momentum and direction with 0.15~GeV/$c$ and 16~$\mu$rad resolutions, respectively. 

For the present analysis, kaon decays in a 50~m long fiducial volume are reconstructed. This volume is contained in a decay tank evacuated to $10^{-6}$~mbar. The momentum and position of the daughter particles are measured by a spectrometer consisting of two straw-tube chambers (STRAW) on either side of a dipole magnet providing a transverse horizontal momentum kick of 270~MeV/$c$. Reconstructed STRAW tracks measure the momentum with a resolution $\sigma_p / p$ in the range of 0.3--0.4\%. Daughter photons are detected by a hermetic system involving two lead-scintillator calorimeters (IRC and SAC) for emission angles with respect to the $Z$ axis $\theta<1$~mrad, a liquid krypton electromagnetic calorimeter (LKr) for $1 < \theta < 10$~mrad, and a system of 12 annular lead-glass detectors (LAV) for $10 < \theta < 50$~mrad.
The detection inefficiency is below $10^{-3}$ for photons directed towards the IRC and SAC calorimeters with energy above 6~GeV; below $10^{-5}$ for photons hitting the LKr calorimeter with energy above 10~GeV; below $10^{-3}$ for photons hitting the LAV detector with energy above 1~GeV. A localized set of LKr cells with coincident signals is grouped into a cluster, providing measurements of energy, transverse coordinates, and time with resolutions of $\sigma_E / E = 4.8\%/\sqrt{E[\mathrm{GeV}]}\oplus 11\%/E[\mathrm{GeV}]\oplus0.9\%$, 1~mm, and between 0.5 and 1~ns depending on the amount and type of energy deposition, respectively. 

A ring-imaging Cherenkov detector (RICH), with 70~ps resolution, identifies secondary charged pions. Two downstream scintillator hodoscopes provide fast time response for charged particles: the CHOD, a matrix of tiles read out by SiPMs, has a time resolution below 1~ns; the NA48-CHOD, composed of two orthogonal planes of scintillator slabs, has 200~ps resolution for coincidence between vertical and horizontal slabs (NA48-CHOD candidate). Two hadronic iron/scintillator-strip sampling calorimeters (MUV1,2) and an array of scintillator tiles located behind 80 cm of iron (MUV3,  with 400~ps time resolution) supplement the pion/muon identification system. The overall probability for identifying a $\mu^+$ as a $\pi^+$ in the momentum range 15--35~GeV$/c$ is at the level of $10^{-7}$~\cite{pnnPaper}.

Information from the NA48-CHOD, CHOD, RICH, MUV3, LKr, and the most downstream LAV station (LAV12) is hardware-processed to issue level-zero (L0) trigger signals with a frequency up to
 1~MHz. The L0 trigger condition used to search for the decay chain of Eq.~\ref{eq:decaychain}, denoted as signal trigger, aims to select final states with one emitted $\pi^+$
and missing energy. It requires a signal in the RICH in coincidence within 10~ns with a signal in at least one CHOD tile. No signals 
in  opposite CHOD quadrants must be found within the 10 ns window, thus reducing the contribution of $K^+\to\pi^+\pi^+\pi^-$ decays and in general of final states with multiple charged particles; this condition is called $QX$-veto in the following. No signals in the MUV3 detector must be present, thus reducing the contribution of $K^+\to(\pi^{0})\mu^+\nu$ decays. No more than one in-time signal must be found in LAV12 and no more than 20~GeV of total energy deposit in time in the LKr calorimeter must be reconstructed. These conditions reduce the contribution of multi-photon final states and are particularly effective in rejecting forward-emitted photons from $\pi^0$ decays.

A software trigger (L1) reconstructs data from the KTAG, LAV and
STRAW detectors to further enforce the presence of a charged kaon and to reject final states with additional particles emitted at large angle. 
The charged kaon must be positively identified using KTAG information within 10~ns of the L0 trigger RICH-based time. At least one STRAW track must be reconstructed, corresponding to a particle with momentum below 50 GeV/$c$ and a point of closest approach (less than 20~cm) to the nominal beam axis upstream of the first STRAW chamber. Events with in-time signals in three or more LAV blocks are rejected.  
These conditions reduce the trigger rate by a factor of 100.

For normalization, the analysis uses data taken with a concurrent minimum-bias L0 trigger (``control trigger'') based on NA48-CHOD information. The control trigger requires one or more time coincidences between horizontal and vertical planes of scintillators in the NA48-CHOD hodoscope, and is downscaled by a factor of 400.

\section{Analysis principle}
Assuming a dominant invisible decay of the $A^\prime$ (or a long-lived $A^\prime$ producing no observable interaction in the LKr calorimeter), the experimental signature for the events described in Eq.~\ref{eq:decaychain} is given by a kaon decaying into a charged pion and a photon hitting the LKr calorimeter, with missing energy and momentum. The kaon and pion momenta are measured with the GTK and STRAW detectors, respectively, and the corresponding 4-momenta are denoted $P_K$ and $P_\pi$. The measurement of the position of impact and the  energy released in the LKr allow the determination of the photon 4-momentum $P_\gamma$, assuming emission from the decay vertex. The squared missing mass
\begin{equation}
M_\mathrm{miss}^2 = \left(P_K - P_\pi - P_\gamma\right)^2
\label{eq:mmiss}
\end{equation}
is expected to peak at $M_{A^\prime}^2$ for the decay chain in Eq.~\ref{eq:decaychain} and at zero for the most abundant background, $\pi^0\to\gamma\gamma$ with one photon undetected. 

A high-purity kinematic identification of the $K^+\to\pi^+\pi^0$ decays is performed by reconstructing solely the $K^+$ and $\pi^+$ particles. The number of $K^+\to\pi^+\pi^0$ decays, 
$n_{\pi^0}$, counted in the control-trigger sample defines the statistics of tagged $\pi^0$ mesons used for normalization.

\begin{figure}[!ht]
  \begin{center}
    \includegraphics[width=12cm]{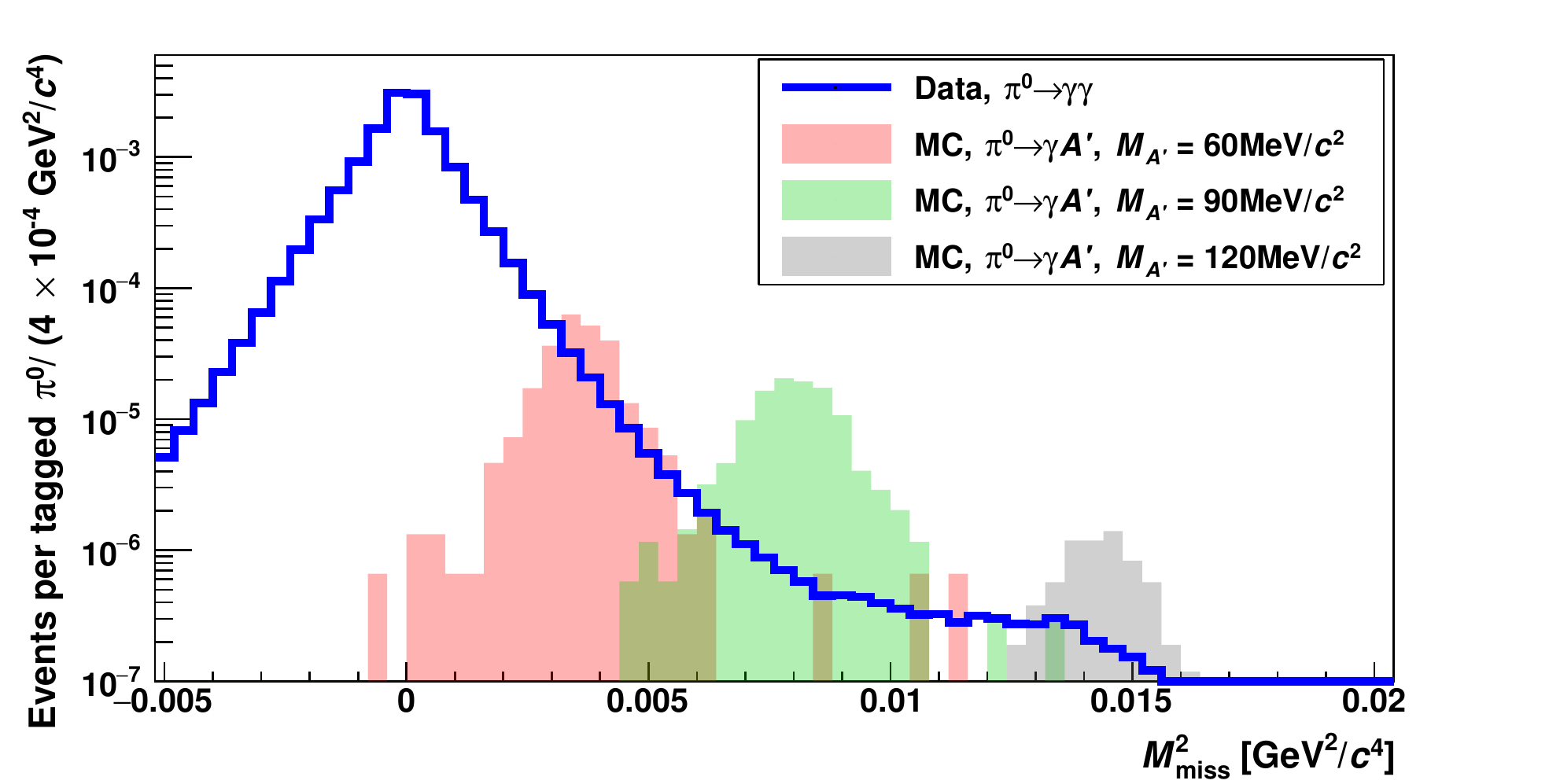}
    \caption{Distributions of the squared missing mass evaluated from $K^+$ decays with one photon and one $\pi^+$ reconstructed (Eq.~\ref{eq:mmiss}). Data from $\pi^0\to\gamma\gamma$ with one photon, randomly chosen, assumed to be undetected are shown by the blue line. The expected spectra from MC simulations of $\pi^0\to A^\prime\gamma$ with a coupling strength $\epsilon^2=2.5\times10^{-4}$ and $A^\prime$ masses of 60 (red), 90 (green) and 120 MeV/$c^2$ (grey) are also shown. For details about the normalization used, see text. }
    \label{fig:CompareDataVsMC}
  \end{center}
\end{figure}

Additional conditions are required for signal-triggered events, in order to enforce the sole presence of a $\pi^+$ and one photon in the final state. The selection efficiency for these additional requirements and the signal-trigger efficiency depend on $M_{A^\prime}$ and are denoted as $\varepsilon_\mathrm{sel}$ and $\varepsilon_\mathrm{trg}$. A peak search in the positive tail of the $M^2_\mathrm{miss}$ background distribution is performed by comparing the number of events in a sliding $M_\mathrm{miss}^2$ window to the background expectation. For illustration, the distributions of $M^2_\mathrm{miss}$ from a Monte Carlo (MC) simulation of the NA62 apparatus when injecting $A^\prime$ signals with masses of 60, 90, and 120~MeV$/c^2$ and a coupling strength $\epsilon^2=2.5\times10^{-4}$ (see Eq.~\ref{eq:br}) are shown in Fig.~\ref{fig:CompareDataVsMC}. These are superimposed on the expected contribution from a control-trigger data sample with fully reconstructed $\pi^0\to\gamma\gamma$
in which one of the two photon LKr clusters, randomly chosen, is artificially excluded. The data distribution is scaled to $n_{\pi^0}$. Each MC distribution is scaled to the equivalent number of tagged $\pi^0$ mesons corresponding to the generated statistics.  

The estimate of the number of signal events $n_\mathrm{sig}$ in a given $M^2_\mathrm{miss}$ window is normalized to the number $n_{\pi^0}$ to yield the BR for the decay $\pi^0\to A^\prime\gamma$ (and hence the $\epsilon^2$ coupling strength):
\begin{equation}
\label{eq:ul}
\mathrm{BR}(\pi^0\to A^\prime\gamma) = \mathrm{BR}(\pi^0\to \gamma\gamma)
\frac{n_\mathrm{sig}}{n_{\pi^0}}\frac{1}{ \varepsilon_\mathrm{sel}\varepsilon_\mathrm{trg}\varepsilon_\mathrm{mass}},
\end{equation}
where the correction factor $\varepsilon_\mathrm{mass}$ accounts for the acceptance of the sliding $M^2_\mathrm{miss}$ window used. 
The geometrical acceptance and the $\pi^0$-tagging efficiency are identical for the signal and normalization channels and therefore cancel exactly in Eq.~\ref{eq:ul}. Part of the sample is solely used for a data-driven background evaluation, reducing the size of the dataset exploited in the signal search.
\subsection{Selection of the normalization sample}
The normalization sample is selected as follows.
\begin{itemize}
\item Events with one charged daughter particle are required: exactly one STRAW good-quality track geometrically associated with a NA48-CHOD candidate must be reconstructed. The STRAW track is the $\pi^+$ candidate in the decay $K^+\to\pi^+\pi^0$. The track momentum must lie in the range $15 < p_\pi < 35~\mathrm{GeV/}c$, ensuring at least 40~GeV of missing energy for a nominal kaon momentum of 75~GeV$/c$.
\item To achieve high-purity pion identification, the $\pi^+$-candidate track is associated in time with a single ring from the RICH consistent with the track direction. The NA48-CHOD hodoscope is used as the time reference for the association to the RICH, which is then used as a reference for all subsequent associations. The track must be geometrically associated with in-time energy deposits from the LKr, MUV1, and MUV2 calorimeters. No in-time MUV3 signal must be geometrically associated with the track. Information from the LKr,  MUV1, and MUV2 is combined in a multivariate classifier 
leading to a $\mu$-to-$\pi$ mis-identification probability of
$10^{-7}$~\cite{pnnPaper}.
\item The $\pi^+$-candidate track must be associated in space and time with exactly one beam track reconstructed with the GTK detector. Tight requirements are applied for this association to minimize the kinematic tails in the reconstruction: the matching time difference must be less than 400 ps and the spatial distance of minimum approach cannot exceed 5 mm. The point of closest approach of the $\pi^+$-candidate and beam tracks is taken as the reconstructed decay vertex. Its  
longitudinal position must lie in the interval $115 < Z < 165~\mathrm{m}$ (Fig.~\ref{fig:NA62Apparatus}). 
\item The beam particle is identified as a charged kaon by its association in time with a KTAG candidate with signals in five or more sectors. The kaon-candidate momentum reconstructed by the GTK spectrometer must lie in the range $72 < p_K < 78~\mathrm{GeV/}c$, to be consistent with the beam momentum.
\item The squared missing mass is required to be consistent with the squared $\pi^0$ mass:
$0.013 < \left(P_K - P_\pi\right)^2 < 0.023~\mathrm{GeV}^2/c^4$.
\end{itemize}
These conditions select 
$K^+\to\pi^+\pi^0(\gamma)$ decays (inclusive of the inner-brems\-strah\-lung radiative component, IB) with contamination below the per-mil level.
The total number of selected events is $1\,030\,155$. After accounting for the control-trigger downscaling factor of 400, the number of tagged $\pi^0$ mesons corresponding to the signal trigger sample is $n_{\pi^0} = 4.12 \times 10^8$. It has been checked that the statistical error on the downscaling has a negligible impact.

\subsection{Selection of the signal sample}
\label{sec:signalSelection}
The algorithm described in the previous section is also applied to signal-triggered events. Further requirements are applied to identify the decay chain of Eq.~\ref{eq:decaychain}.
\begin{itemize}
\item No in-time signals from the LAV and SAC-IRC systems must be present.
\item Exactly one in-time LKr cluster with energy $E_\gamma> 1.5$~GeV is required at least 20~cm away from the pion impact point. The selected LKr cluster is assumed to be due to a photon originating from the decay vertex: its energy and position are used to evaluate the photon momentum. The missing momentum $\vec{p}_\mathrm{miss}$ evaluated from the kaon, pion, and photon momenta must extrapolate from the decay vertex to the LKr calorimeter and any activity in the LKr around the $p_\mathrm{miss}$ impact point must not have a total energy in excess of 1~GeV. These conditions ensure further rejection against additional photons. The impact point of $\vec{p}_\mathrm{miss}$ must be
at least 20 cm away from the LKr clusters associated with the pion and photon, thus minimizing energy sharing (isolation cut).

\item No in-time RICH signals may be found apart from those reconstructing the pion Cherenkov ring, thus minimizing the contribution from upstream photon conversions in the STRAW chamber and RICH vessel materials.

\item A reconstruction bias may occur when a photon converts before reaching the LKr sensitive volume: if one particle of the $e^+e^-$ pair from the conversion is undetected, the energy of the reconstructed photon cluster tends to be underestimated, occasionally by several GeV. This effect has an impact on the background due to $\pi^0\to\gamma\gamma$ decays with one photon lost. The energy of the undetected photon is usually below 1~GeV, therefore a bias in the reconstruction of the detected photon may induce a correlated shift of the missing energy and of $M_\mathrm{miss}^2$ towards positive values. 
Moreover, events with a systematic underestimation of the detected photon energy may have the impact point of $\vec{p}_\mathrm{miss}$ in the LKr sensitive region, whereas the missing photon truly points to the LAV system. Imposing a lower threshold to the missing energy mitigates these effects, as shown by MC simulation. The missing energy evaluated from the energies of the kaon, pion, and photon LKr cluster, $E_\mathrm{miss}=E_{K} - E_\pi - E_{\gamma}$, is required to be at least 5~GeV above its kinematic lower limit, calculated for the decay of a $\pi^0$ to a photon and a particle of mass squared $M^2_\mathrm{miss}$.

\item No in-time NA48-CHOD candidates must be found except for those geometrically associated with the $\pi^+$.
This condition is referred to as the \textit{NA48-CHOD Extra-activity cut}.
\end{itemize}
A total of $8\,915$ events satisfy these criteria.
\subsection{Background evaluation}
\label{sec:backgroundEvaluation}
After signal selection, MC studies suggest that all background events are $K^+\to\pi^+\pi^0(\gamma)$ decays, in which one of the photons from the decay $\pi^0\to\gamma\gamma$ is lost due to photo-nuclear interactions or conversions downstream of the NA48-CHOD hodoscope. The detected photon might be correctly reconstructed or its energy might be systematically underestimated due to conversion downstream of the NA48-CHOD hodoscope (e.g. in the LKr cryostat). 
Background channels $\kenp , \kmnp , \kppgDE , \kppgINT , \kmnug , \kenug $ and $\kppp$ are expected to yield less than one selected event (DE refers to the direct emission component, while INT refers to the interference between DE and IB amplitudes).

\begin{figure}
\begin{center}
\includegraphics[width=8.2cm]{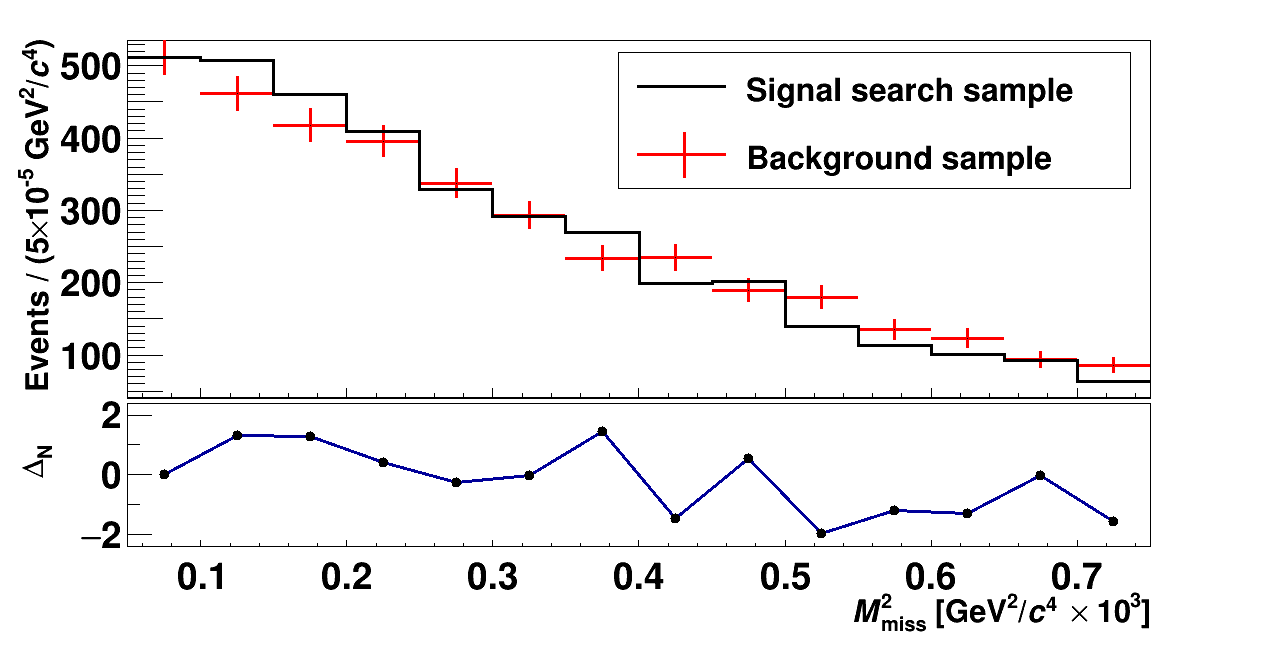}\includegraphics[width=8.2cm]{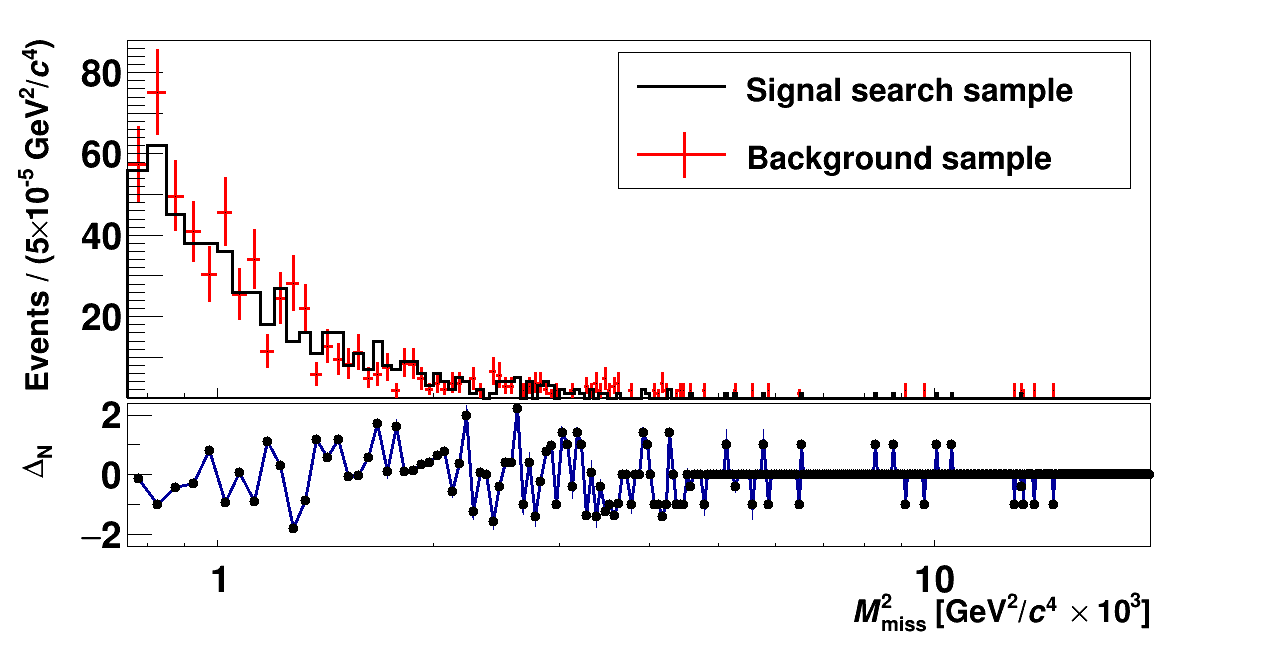}
\end{center}
\caption{$M^2_\mathrm{miss}$ distribution of samples for $A^\prime$ search (black) and background (red, with error bars). The scaling factors are evaluated in the region shown in the left panel. The search region is shown in the right panel. In the bottom panels, the difference $\Delta_N$ between the two $M^2_\mathrm{miss}$ spectra  in units of its standard deviation is shown.}
\label{fig:SpectrumComparison}
\end{figure}

To evaluate the expected background, a data-driven approach is used. The data selection of Sec.~\ref{sec:signalSelection} is applied but the \textit{NA48-CHOD Extra activity cut} is partially inverted: events with one in-time NA48-CHOD candidate geometrically associated with the detected photon are rejected, while the presence of candidates far from both the $\pi^+$ and photon impact points to the NA48-CHOD hodoscope is required. This allows the selection of a data control sample of $\pi^0\to\gamma\gamma$ events with one photon lost because of conversion upstream of the NA48-CHOD. Ensuring the presence of a second photon with no overlap with the signal sample, the control sample can be used to evaluate the expected $M_\mathrm{miss}^2$ background distribution with a bias that is below the statistical uncertainty, as verified by MC simulations. The control sample is scaled to the signal sample in a side-band region adjacent to but not overlapping with the $A^\prime$ search region. Background considerations suggest considering a minimum mass of 30~MeV$/c^2$ for the $A^\prime$ search. Similarly, acceptance and yield considerations suggest considering a maximum mass of 130~MeV$/c^2$. Given the expected mass resolution discussed in the next section, the search region is $0.00075 < M^2_\mathrm{miss} < 0.01765~\mathrm{GeV}^2/c^4$. The scaling window used corresponds to $0.00005 < M^2_\mathrm{miss} < 0.00075~\mathrm{GeV}^2/c^4$ (Fig.~\ref{fig:SpectrumComparison}, left).

Particular care has been taken to avoid a possible trigger-induced bias when evaluating the expected background. 
The signal trigger applies the $QX$-veto condition, rejecting events with in-time signals in opposite CHOD quadrants. 
To account for the $QX$-veto potential inefficiency, signal-selected and background samples are divided according to whether the impact points of $\vec{p}_\mathrm{miss}$ and of the charged pion track lie in opposite CHOD quadrants or not.
The uncertainty on the scale factors is included in the evaluation of the upper limit. The distributions of $M_\mathrm{miss}^2$ for the signal search and the scaled background samples are shown in  Fig.~\ref{fig:SpectrumComparison}, right.

\section{Search for an $\bm{A^\prime}$ signal}
\label{Asearch}
The expected $M_\mathrm{miss}^2$ distribution for an $A^\prime$ signal and the selection efficiency are evaluated using MC simulations of the $\pi^0\to A^\prime\gamma$ decay, with $M_{A^\prime}$ ranging from 30 to 130~MeV/$c^2$ in steps of 10 MeV/$c^2$.

\begin{figure}[!ht]
\begin{center}
\includegraphics[width=12cm]{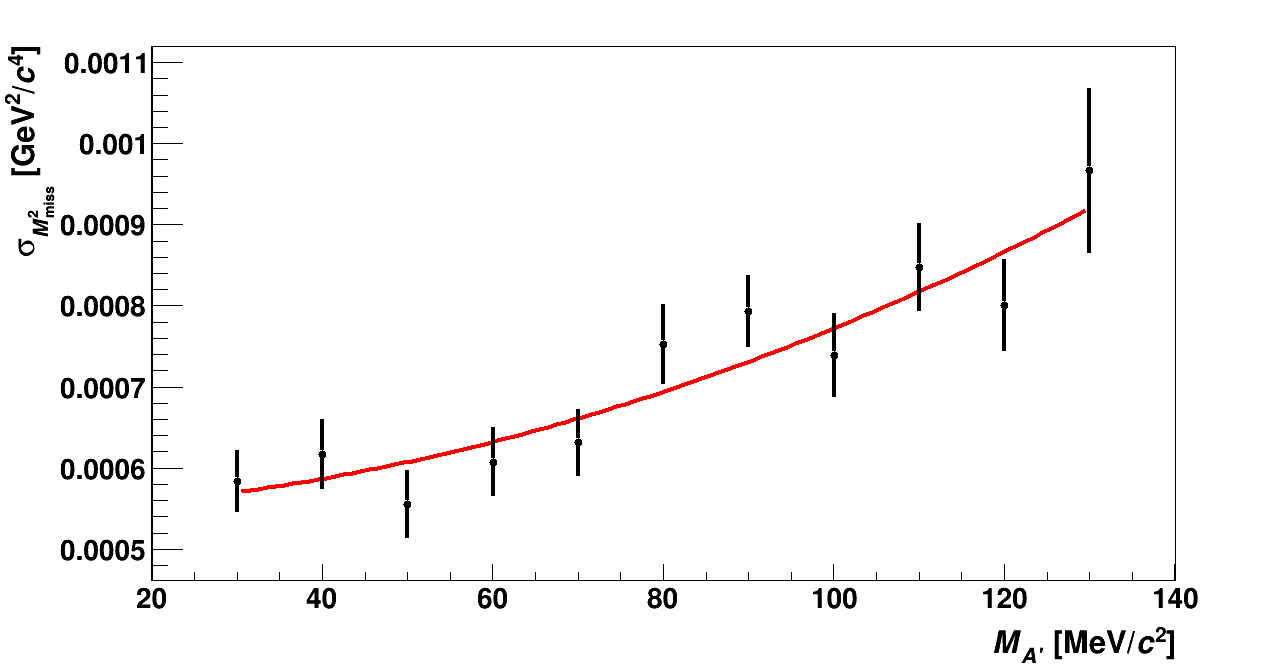}
\end{center}
\caption{Resolution on the squared missing mass for events satisfying the signal selection criteria, 
from the simulation of the decay chain $K^+ \to \pi^+\pi^0$ followed by $\pi^0 \to A^\prime \gamma$, as a function of $M_{A^\prime}$. A polynomial function describing the mass dependence is also displayed.}
\label{fig:FitMCMasses}
\end{figure}

Given the expected background, for each $A^\prime$ mass value the signal region optimizing the upper limit in a background-only hypothesis is defined as a $\pm 1\,\sigma_{M^2_\mathrm{miss}}$ window around the expected $M^2_\mathrm{miss}$ peak value, where $\sigma_{M^2_\mathrm{miss}}$ is the resolution. 
The resolution slowly degrades with increasing $M_{A^\prime}$ (Fig.~\ref{fig:FitMCMasses}). This behaviour is dominated by the relative resolution on the photon energy measured with the LKr calorimeter: the higher the $A^\prime$ mass, the lower the energy of the detected photon. 
The dependence of the resolution on the mass is parameterized with a polynomial function to allow interpolation in the whole search region.

To determine the reliability of the simulation of the missing mass resolution,
data and MC simulation are compared for fully reconstructed $\kpp$, $\pgee$ decays. The resolution on $\left(P_K - P_\pi - P_\gamma \right)^2 - \left(P_{e^+} + P_{e^-} \right)^2$ is studied as a function of the di-lepton mass ($P_{e^+}$ and $P_{e^-}$ are the positron and electron 4-momenta). 
Data and MC resolutions are found to agree within 10\%. The uncertainty on the resolution is considered in the evaluation of the systematic error on the observed limit.

\subsection{Efficiency corrections}

The selection efficiency, $\varepsilon_\mathrm{sel}$, is evaluated by MC simulation.
A study of the signal loss due to effects not included or not reliably modeled in the MC simulation has been performed. The signal selection requires the presence of a single photon in the final state. 
Therefore, an efficiency loss is expected due to in-time accidental activity from upstream decays of kaons and pions in the beam, or to the decay $K^+\to\pi^+\pi^0\gamma$ if the radiative photon is sufficiently hard to be detected. 
The expected contribution from the former source, the so-called random-veto effect, is evaluated with data collected with the control trigger: information from out-of-time windows from various data samples of $K^+\rightarrow \pi^+ \pi^0$, $\pi^0\rightarrow \gamma \gamma$ decays is used for this purpose. For the latter source, MC simulations of the radiative decay are combined with the measured photon detection efficiency. The overall loss due to the two effects is $(19.7\pm0.2_\mathrm{stat}\pm1.5_\mathrm{syst})\%$,
 dominated by the random-veto contribution. The systematic error includes an estimate of the reliability of the control samples used to reproduce the random-veto effect for the signal sample (0.7\%) and a conservative evaluation of the uncertainty on the detection efficiency of the radiative photon (1.3\%).

The trigger efficiency, $\varepsilon_\mathrm{trg}$, is evaluated using an MC simulation with data inputs. Data samples of $K^+ \to \pi^+\pi^0$ events are selected from the control trigger in which exactly one photon LKr cluster is present. The missing momentum $\vec{p}_\mathrm{miss}$ must point towards one of the LAV stations, thus ensuring the absence of the second photon in the LKr calorimeter. To mimic the signal sample, in which the $QX$-veto condition is applied, this control sample is reduced to geometrical configurations in which the selected photon and the charged pion do not traverse opposite CHOD quadrants. The signal-trigger efficiency is obtained as the fraction of events satisfying the signal-trigger chain. The efficiency is binned as a function of the total energy released in the LKr calorimeter. To reproduce the trigger condition, which requires the total energy release to be below 20~GeV, the binned efficiency is used as an event-weight for the MC simulation of the signal. The expected distribution of the total LKr energy for the signal is therefore convoluted with the data-measured trigger efficiency. The inefficiency induced by the L1 is determined with data-driven methods and is found to be less than 3\%.

\begin{figure}[!ht]
\begin{center}
\includegraphics[width=12cm]{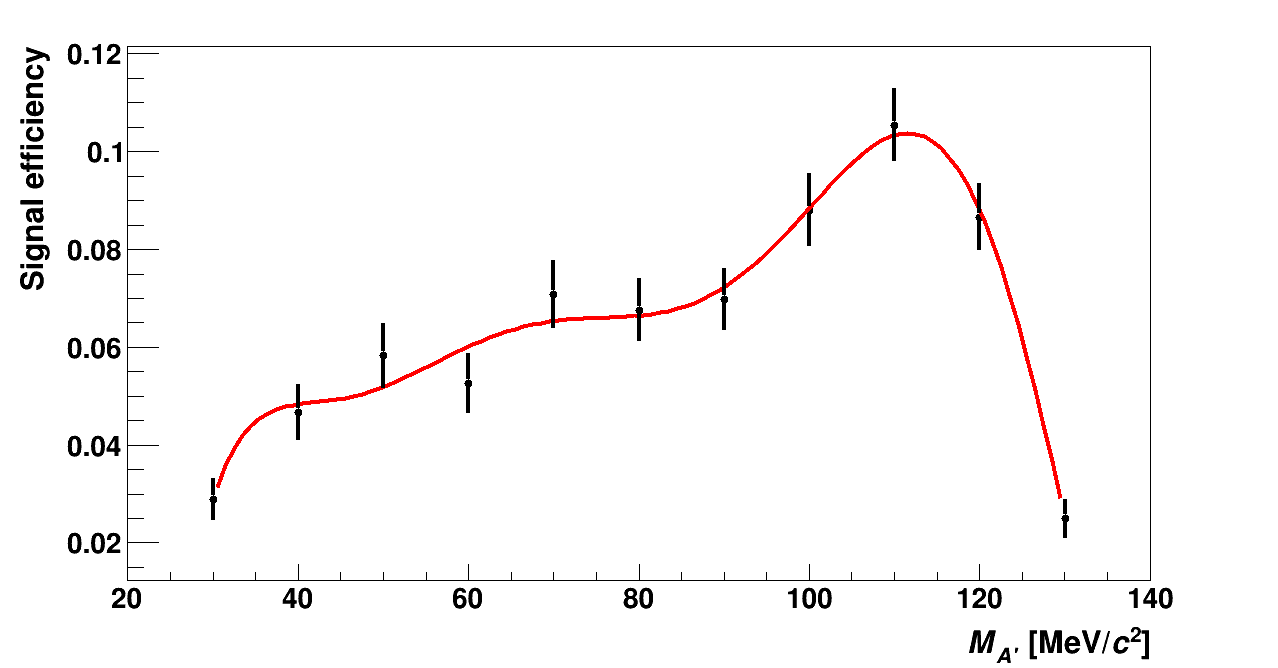}
\end{center}
\caption{Total efficiency as a function of $M_{A^\prime}$. A polynomial function is used to interpolate the global efficiency in the range $30<M_{A^\prime}<130$~MeV/$c^2$.}
\label{fig:AprimeEff}
\end{figure}

The total efficiency combining $\varepsilon_\mathrm{sel}$, $\varepsilon_\mathrm{trg}$, and the mass-window acceptance $\varepsilon_\mathrm{mass}$ determined by MC simulation is shown as a function of $M_{A^\prime}$ in Fig.~\ref{fig:AprimeEff}.
It is  parameterized with a polynomial function to interpolate in the range 30~MeV/$c^2 < M_{A^\prime} < 130$~MeV/$c^2$. The dependence of the efficiency on $M_{A^\prime}$ is dominated by kinematic effects: a heavy $A^\prime$ is emitted collinear to a soft visible photon, thus losses can occur both due to the photon detection efficiency and to the isolation cut.

\subsection{Evaluation of the upper limit}
\label{uleval}
The observed data and the expected background counts are evaluated by integrating the corresponding $M^2_\mathrm{miss}$ spectrum (Fig.~\ref{fig:SpectrumComparison}, right) in a $\pm1\,\sigma_{M^2_\mathrm{miss}}$ signal search window. To avoid the case of exactly zero expected counts, background events lying above 0.005~$\mathrm{GeV^2}/c^4$ (``flat region'') are grouped into a single bin. For the signal windows that overlap the flat region, the background entries in the single bin are scaled by the ratio of the signal window width to the flat region width and the errors are evaluated accordingly.

Using the CL$_\mathrm{s}$ algorithm~\cite{CLsArticle}, frequentist 90\% confidence intervals are determined for the number of signal events. The upper limits are compatible within two standard deviations with the fluctuation expected in a background-only scenario.

\begin{figure}[!ht]
\begin{center}
\includegraphics[width=12cm]{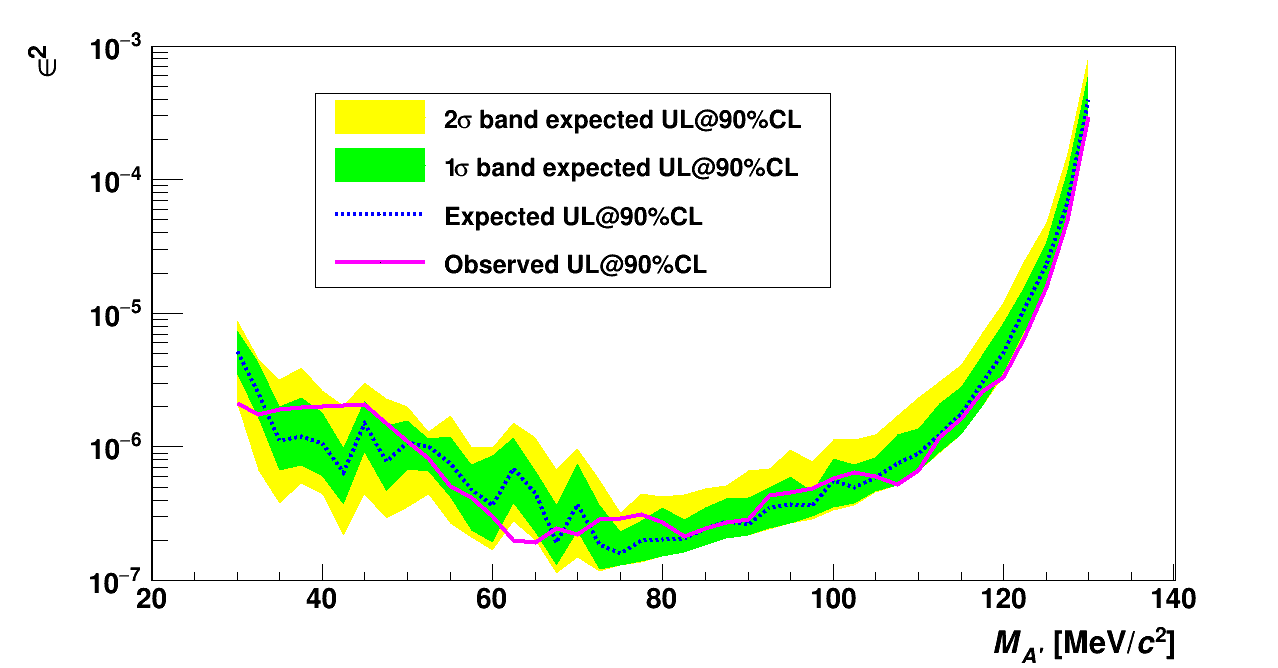}
\end{center}
\caption{Upper limits at 90\% CL on the dark photon coupling strength ($\epsilon^2$) as a function of the mass ($M_{A^\prime}$). The limit obtained from data (solid line) should be compared to that expected in the absence of signal: the median of the upper-limit distribution in the background-only hypothesis is shown by the dashed line and the corresponding fluctuation bands with 68\% and 95\% coverage are shown by the shaded areas.}
\label{fig:uls}
\end{figure}

The 90\% CL upper limits obtained on the coupling parameter
$\epsilon^2$ as a function of $M_{A^\prime}$ are shown in Fig.~\ref{fig:uls}. The limit from the number of observed events (solid curve) is compared to the bands with 68\% and 95\% coverage in the absence of signal: no statistically significant excess is detected.

\subsection{Systematic errors}
\label{sec:syst}
Various parameters used in the statistical procedure have been varied to evaluate the systematic uncertainty on the calculated upper limits. The lower edge of the window used to evaluate the scale factors to compare background and signal-search samples has been varied using the following additional values: $-0.00015$, 0.00015, 0.00025~GeV$^2/c^4$. The first value implies using the peak of the background distribution for scaling, while the other values correspond to using smaller and smaller portions towards positive values of the $M^2_\mathrm{miss}$ distribution. For each of these choices, the scaling has been re-evaluated and an upper limit has been obtained.
The signal window has been varied to $\pm 0.9$, $\pm 1.1$, and $\pm 2\,\sigma_{M^2_\mathrm{miss}}$.
The extent of the flat region has been varied by moving its lower edge to 0.004 and 0.006~$\mathrm{GeV^2/}c^4$: these two values correspond to a variation larger than one standard deviation of the signal distribution.

The uncertainties on the signal efficiency, including statistical and systematic errors, have been considered in the evaluation of the upper limit. A confidence band has been calculated for the polynomial interpolation based on the ten efficiency points of Fig.~\ref{fig:AprimeEff}: each interpolated value is taken with the total relative uncertainty of the nearest efficiency point. Moreover, for each efficiency point, the results obtained using the polynomial interpolation have been compared to those using the central values. 

No significant deviation beyond the statistical uncertainty has been observed in these studies.

To prove the discovery sensitivity of the analysis, a dark photon signal is injected into the data and the statistical treatment is applied to this altered sample. The $M^2_\mathrm{miss}$ spectrum corresponding to the MC simulation of an $A^\prime$ with 80~MeV/$c^2$ mass is scaled according to four different values of the coupling strength $\epsilon^2$: $6.4\times 10^{-7}$, $10^{-6}$, and $4\times 10^{-6}$. The scale factor applied to each $A^\prime$ signal takes into account the full selection and trigger efficiency with its uncertainty. The scaled histograms are added to the data distribution. 

The upper limits for these altered samples demonstrate that the method described in this work is able to detect such $A^\prime$ signals: for all of the above $\epsilon^2$ values, the upper limits found exceed the limit from the background-only hypothesis beyond its 95\% coverage.

\section{Search for the $\bm{\pgnn}$ decay}
With slight modifications to the analysis, a search has been conducted for the decay $\pgnn$, for which the BR is expected to be of the order of $10^{-18}$~\cite{wolfram} within the SM.
The present experimental limit is BR($\pgnn$)$<6\times 10^{-4}$ at 90\% CL~\cite{Atiya92}. The strategy to search for this decay is the same as that used for the $A^\prime$, based on the comparison of data and expected background counts in a given $M^2_\mathrm{miss}$ interval. A MC simulation of the decay is performed using the phase-space density in~\cite{wolfram}. The combined efficiency for the signal selection of Sec.~\ref{sec:signalSelection} and the trigger conditions is $(14.0\pm0.6)\%$. The range $0.0054\mbox{ GeV}^2/c^4 < M^2_\mathrm{miss} < M^2_{\pi^0}$ is used as the signal-search window after MC optimization of the expected limit in the background-only hypothesis. The $M^2_\mathrm{miss}$ window chosen contains $(52.8\pm 1.7)\%$ of the signal events. 
The number of observed events is 7, while the expected background counts are $12\pm4$.
The CL$_\mathrm{s}$ technique provides an upper limit on the number $N_{s}$ of decays observed, $N_s < 5.6$, which is compatible (within one standard deviation) with the results from the expected background fluctuations. After applying the efficiency corrections, the 90\% CL upper limit obtained is:
\begin{equation}
\label{eq:wolframRes}
\mathrm{BR}(\pi^0\to \gamma\nu\bar{\nu}) < 1.9 \times 10^{-7}.
\end{equation}

\section{Conclusions}
\label{summary}
\begin{figure}[ht]
\begin{center}
\includegraphics[width=13cm]{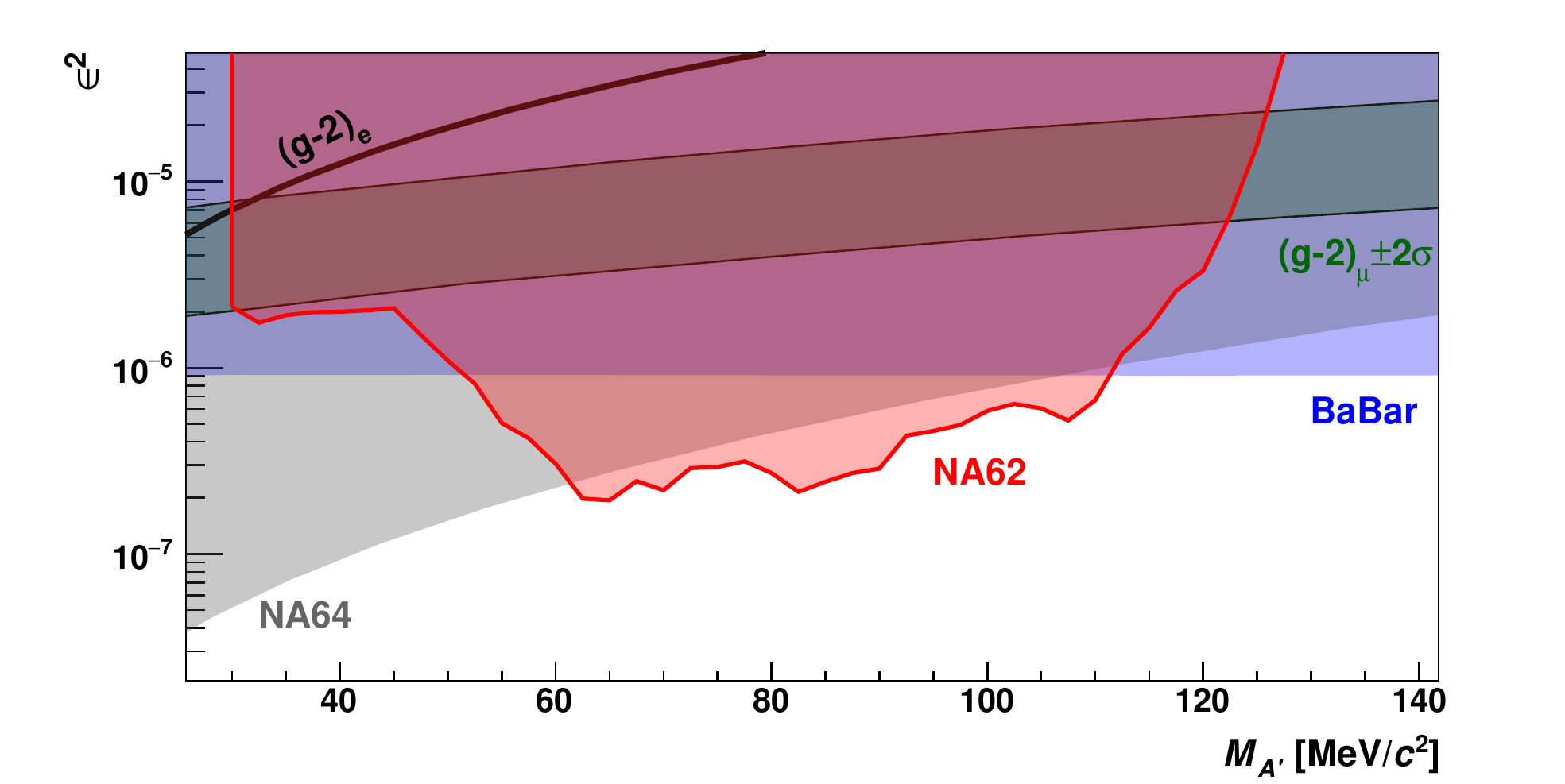}
\end{center}
\caption{
Upper limit at 90\% CL from NA62 (red region) in the $\epsilon^2$ vs $M_{A^\prime}$ plane with $A^\prime$ decaying into invisible final states.
The limits from the BaBar~\cite{babar} (blue) and NA64~\cite{NA64} (light grey) experiments are shown. The green band shows the region of the parameter space corresponding to an explanation of the discrepancy between the measured~\cite{Bennett:2006fi} and expected values of the anomalous muon magnetic moment $(g-2)_{\mu}$~\cite{amu} in terms of a contribution from the $A^\prime$ in the quantum loops~\cite{darkphotonForAmu1,darkphotonForAmu2}. The region above the black line is excluded by the agreement of the anomalous magnetic moment of the electron $(g-2)_{e}$ with its expected value~\cite{g_2e_1,g_2e_2,g_2e_3}.}
\label{fig:Comparison}
\end{figure}

A search for an invisible dark photon $A^\prime$ has been performed, exploiting the efficient photon-veto capability and high resolution tracking of the NA62 detector. The signal stems from the chain $K^+ \to \pi^+\pi^0$ followed by $\pi^0 \to A^\prime \gamma$. Given the kaon, charged pion, and photon 4-momenta, the squared missing mass $M_\mathrm{miss}^2 = \left(P_K - P_\pi - P_\gamma\right)^2$ is expected to peak at the squared $A^\prime$ mass for the signal and at zero for the dominant background, $\pi^0\to\gamma\gamma$ decays with one photon undetected. A peak search has been conducted, comparing signal-selected samples and data-driven background estimates. Using the CL$_\mathrm{s}$ method, no significant statistical excess has been identified and upper limits on the coupling strength $\epsilon^2$ in the mass range 30--130 MeV/$c^2$ have been set,
improving  on the previous limits over the mass range 60--110 MeV/$c^2$ (Fig.~\ref{fig:Comparison}).

It should be noted that the experimental technique used here differs from that of previous results. At BaBar, positron-electron annihilations to one photon and one dark photon at the centre of mass energy of the $\Upsilon$ resonances should produce energetic single-photon events~\cite{babar}. At NA64, dark photons produced by a 100 GeV electron beam dumped into a calorimeter are supposed to yield an excess of events with large missing energy~\cite{NA64}. As a consequence, models different from that of Eq.~\ref{eq:mixing} and e.g. involving suppressed dark-photon lepton couplings~\cite{serendipity}, might produce a signal at NA62 notwithstanding the NA64 and BaBar experimental results.
The measurement of the BR for the decay $K^+\to\pi^+\nu\overline{\nu}$ by the E787 and E949 experiments~\cite{brookheven} can be interpreted as a limit on the BR for the decay $K^+\to\pi^+A^\prime$ as a function of the $A^\prime$ mass. However, this interpretation is model-dependent: if a mixing of the dark photon to the $Z$ boson is introduced, the lower edge of the exclusion bound increases by a factor of 7~\cite{davoudiasl2014}. In the most conservative scenario, not shown in Fig.~\ref{fig:Comparison}, the upper limit from E787-E949 partially overlaps with the $(g-2)_\mu$ band in the mass ranges 83--113 and 176--243 MeV$/c^2$. 

Finally, an upper limit has been set for the branching ratio of the decay $\pi^0 \to \gamma \nu \bar\nu$, $\mathrm{BR} < 1.9 \times 10^{-7}$ at 90\% CL, improving the current limit by more than three orders of magnitude.
\section*{Acknowledgements}
It is a pleasure to express our appreciation to the staff of the CERN laboratory and the technical
staff of the participating laboratories and universities for their efforts in the operation of the
experiment and data processing.

The cost of the experiment and of its auxiliary systems were supported by the funding agencies of 
the Collaboration Institutes. We are particularly indebted to: 
F.R.S.-FNRS (Fonds de la Recherche Scientifique - FNRS), Belgium;
BMES (Ministry of Education, Youth and Science), Bulgaria;
NSERC (Natural Sciences and Engineering Research Council), Canada;
NRC (National Research Council) contribution to TRIUMF, Canada;
MEYS (Ministry of Education, Youth and Sports),  Czech Republic;
BMBF (Bundesministerium f\"{u}r Bildung und Forschung) contracts 05H12UM5, 05H15UMCNA and 05H18UMCNA, Germany;
INFN  (Istituto Nazionale di Fisica Nucleare),  Italy;
MIUR (Ministero dell'Istruzione, dell'Universit\`a e della Ricerca),  Italy;
CONACyT  (Consejo Nacional de Ciencia y Tecnolog\'{i}a),  Mexico;
IFA (Institute of Atomic Physics),  Romania;
INR-RAS (Institute for Nuclear Research of the Russian Academy of Sciences), Moscow, Russia; 
JINR (Joint Institute for Nuclear Research), Dubna, Russia; 
NRC (National Research Center)  ``Kurchatov Institute'' and MESRF (Ministry of Education and Science of the Russian Federation), Russia; 
MESRS  (Ministry of Education, Science, Research and Sport), Slovakia; 
CERN (European Organization for Nuclear Research), Switzerland; 
STFC (Science and Technology Facilities Council), United Kingdom;
NSF (National Science Foundation) Award Number 1506088,   U.S.A.;
ERC (European Research Council)  ``UniversaLepto'' advanced grant 268062, ``KaonLepton'' starting grant 336581, Europe.

Individuals have received support from:
Charles University (project GA UK number 404716), Czech Republic;
Ministry of Education, Universities and Research (MIUR  ``Futuro in ricerca 2012''  grant RBFR12JF2Z, Project GAP), Italy;
Russian Foundation for Basic Research  (RFBR grants 18-32-00072, 18-32-00245), Russia; 
the Royal Society  (grants UF100308, UF0758946), United Kingdom;
STFC (Rutherford fellowships ST/J00412X/1, ST/M005798/1), United Kingdom;
ERC (grants 268062,  336581 and  starting grant 802836 ``AxScale'').


\clearpage



\begin{thebibliography}{99}
\bibitem{Okun} 
L. Okun, Sov. Phys. JETP 56 (1982) 502.
%
\bibitem{Holdom} 
B. Holdom, Phys. Lett. B 166 (1986) 196.
%
\bibitem{NA62paper}
E. Cortina Gil, et al. [The NA62 Collaboration], J. Instrum. 12 (2017) P05025.


\bibitem{pnnPaper}
E. Cortina Gil, et al. [The NA62 Collaboration], Phys. Lett. B 791 (2019) 156.



\bibitem{CLsArticle}
A. L. Read, J. Phys. G 28 (2002) 2693.
%

\bibitem{wolfram}
L. Arnellos, W. J. Marciano, and Z. Parsa, Nucl. Phys.
B 196 (1982) 365.

\bibitem{Atiya92}
M.~S.~Atiya, et al. [The E787 Collaboration],
  Phys.\ Rev.\ Lett.\  69 (1992) 733.


\bibitem{babar} 
J.~P.~Lees, et al. [The BaBar Collaboration], Phys. Rev. Lett. 119 (2017) 131804.


\bibitem{NA64} D. Banerjee, et al. [The NA64 Collaboration]
Phys. Rev. D 97 (2018) 072002.


\bibitem{serendipity}
  P.~Ilten, Y.~Soreq, M.~Williams and W.~Xue,
  JHEP 1806 (2018) 004.

\bibitem{Bennett:2006fi} 
  G.~W.~Bennett, et al. [The Muon g-2 Collaboration],
  Phys.\ Rev.\ D 73 (2006) 072003.

\bibitem{amu} ``Muon Anomalous Magnetic Moment'', within~M. Tanabashi, et al. [Particle Data Group], Phys. Rev. D 98 (2018) 030001.

\bibitem{darkphotonForAmu1} P. Fayet, Phys. Rev. D 75 (2007) 115017. 
\bibitem{darkphotonForAmu2} M. Pospelov, Phys. Rev. D 80 (2009) 095002.
\bibitem{g_2e_1} D. Hanneke, S. Fogwell and G. Gabrielse, Phys. Rev. Lett. 100 (2008) 120801.
\bibitem {g_2e_2} R. Bouchendira, et al., Phys. Rev. Lett. 106 (2011) 080801. 
\bibitem{g_2e_3} T. Aoyama, M. Hayakawa, T. Kinoshita and M. Nio, Phys. Rev. Lett. 109 (2012) 111807.

\bibitem{brookheven} A. V. Artamonov, et al., [The E949 Collaboration], Phys. Rev. D 79 (2009) 092004.

\bibitem{davoudiasl2014} H. Davoudiasl, H. S. Lee and W. J. Marciano, Phys. Rev. D 89 (2014) 095006.

\end{thebibliography}
\end{document}